\documentclass[twocolumn,preprintnumbers,amsmath,amssymb,pra]{revtex4}

\usepackage{graphicx}
\usepackage{dcolumn}
\usepackage{bm}
\usepackage{amssymb}
\usepackage{epstopdf}
\usepackage{color}

\begin{document}

\title{Floquet theory of spin dynamics under circularly polarized light pulses}

\author{O. V. Kibis}\email{Oleg.Kibis@nstu.ru}
\affiliation {Department of Applied and Theoretical Physics, Novosibirsk~State~Technical~University,
Karl~Marx~Avenue~20,~Novosibirsk~630073,~Russia}

\begin{abstract}
Within the Floquet theory of periodically driven quantum systems,
the nonlinear single-spin dynamics under pulse of a circularly polarized electromagnetic field is analyzed. It is demonstrated that the field, first, lifts the spin degeneracy and, second, the field-induced spin splitting is accompanied by the photon emission at the spitting frequency. This two-stage process leads, particularly, to the polarization of spins along angular momentum of the circularly polarized field. As a result, the pulse-induced magnetization appears, what can be observed in state-of the-art measurements.
\end{abstract}
\maketitle

\section{Introduction}
Controlling electronic properties by an off-resonant high-frequency electromagnetic field, which is based physically on the Floquet theory of periodically driven quantum systems (Floquet engineering), remains an exciting research area during the last decades~\cite{Oka_2019,Basov_2017,Eckardt_2015,Goldman_2014,Bukov_2015,Casas_2001,Kibis_2020}. Since frequency of the off-resonant field lies far from
resonant frequencies of the electronic system, it cannot be absorbed by electrons and only periodically drives them (``dresses'' them), modifying all electronic characteristics. As a consequence, the dressing field can crucially change physical
properties of various solids, including semiconductor quantum wells~\cite{Lindner_2011,Pervishko_2015,Dini_2016}, quantum rings~\cite{Kibis_2011,Koshelev_2015,Kozin_2018}, topological insulators~\cite{Rechtsman_2013,Wang_2013,Torres_2014,Calvo_2015,Mikami_2016}, carbon nanotubes~\cite{Kibis_2021},
graphene and related two-dimensional materials~\cite{Oka_2009,Kibis_2010,Iurov_2017,Iurov_2013,Syzranov_2013,Usaj_2014,Perez_2014,Glazov_2014,Sentef_2015,Sie_2015,Kibis_2017,Iurov_2019,Iurov_2020,Cavalleri_2020}, etc. Concerning influence of the dressing field on spin properties, it was studied before for two-dimensional electronic systems~\cite{Pervishko_2015}, quantum rings with the Rashba spin-orbit interaction~\cite{Kozin_2018} and spin transistors~\cite{Sheremet_2016}. However, the Floquet theory of spin dynamics under short pulses of the dressing field waits still for detailed analysis. This problem is of current importance due to the effect
of optically induced magnetization in thin films under circularly polarized light pulses~\cite{Stanciu_2007}, which is considered as a much promised method of ultrafast magnetic recording. Since the effect is very important as a basis for information processing in various devices, it attracts enormous interest of the scientific community up to now~\cite{Kirilyuk_2010,Kimel_2015}. However, its physical nature remains still opened for discussion. The present article is aimed to develop the Floquet theory of single spin dynamics under circularly polarized light pulses and consider the light-induced magnetization effect from its point of view.

The article is organized as follows. In Sec.~II, the effective Hamiltonian describing the behaviour of single spin in a circularly polarized dressing field is constructed. In Sec.~III, the Floquet problem with the Hamiltonian is solved for the particular case of the rectangular field pulse and the found solutions of the problem are analyzed. The last two sections contain conclusion and acknowledgments.

\section{Model}
Let us analyze the interaction between an electron spin and a circularly polarized electromagnetic wave (dressing field) with the magnetic field amplitude $H_0$ and the frequency $\omega_0$, which propagates along the $z$ axis and is clockwise-polarized (see Fig.~1a). Then the Hamiltonian of the interaction reads
\begin{equation}\label{H}
\hat{\cal H}_0=\mu_B{\bm{\sigma}}\mathbf{H},
\end{equation}
where $\mathbf{H}$ is the magnetic field of the wave, $\mu_B=|e|\hbar/2m_ec$ is the Bohr magneton, $m_e$ is the electron mass, and $\bm{\sigma}=(\sigma_x,\sigma_y,\sigma_z)$ is the Pauli matrix vector. As a starting point, we consider the case of monochromatic (non-pulsating) field
\begin{equation}\label{H0}
\mathbf{H}=\frac{H_0}{\sqrt{2}}\left[\mathbf{e}_0e^{-i\omega_0t}+\mathbf{e}^\ast_0e^{i\omega_0t}\right],
\end{equation}
where $\mathbf{e}_{0}=(\mathbf{e}_{x}+ i\mathbf{e}_{y})/\sqrt{2}$
is the polarization vector of the field, and $\mathbf{e}_{x,y}$ are the unit vectors directed along the $x,y$ coordinate axes pictured in Fig.~1a. In the most general form, the nonstationary Schr\"odinger equation with the Hamiltonian (\ref{H}) and the periodically time-dependent field (\ref{H0}) can be written as
$i\hbar\partial_t\psi(t)=\hat{\cal H}_0(t)\psi(t)$, where $\hat{\cal
H}_0(t+T)=\hat{\cal H}_0(t)$ and $T=2\pi/\omega_0$ is the field period. It follows
from the well-known Floquet theorem that solution of the Schr\"odinger
equation is the Floquet function, $\psi(t)=e^{-i\varepsilon
t/\hbar}\varphi(t)$, where $\varphi(t+T)=\varphi(t)$ is the
periodically time-dependent function and $\varepsilon$ is the
(quasi)energy~\cite{Oka_2019,Basov_2017,Eckardt_2015,Goldman_2014,Bukov_2015,Casas_2001}.
The Floquet problem is aimed to find the full set of the Floquet functions, $\psi(t)$, and the corresponding energy spectrum,
$\varepsilon$. In the particular case of the Hamiltonian (\ref{H}) with the field (\ref{H0}), the Floquet problem can be solved accurately. As a result, we arrive at the exact Floquet eigenstates of the Hamiltonian (\ref{H}) describing the spin dynamics under irradiation by the field (\ref{H0}),
\begin{align}
&|\psi_g(t)\rangle=\left[\sqrt{\frac{\Omega+\omega_0}{2\Omega}}|\psi_+\rangle
-e^{i\omega_0t}\sqrt{\frac{\Omega-\omega_0}{2\Omega}}|\psi_-\rangle\right]\nonumber\\
&\times e^{-i\varepsilon_gt/\hbar},\label{Fg}\\
&|\psi_e(t)\rangle=\left[\sqrt{\frac{\Omega+\omega_0}{2\Omega}}|\psi_-\rangle
+e^{-i\omega_0t}\sqrt{\frac{\Omega-\omega_0}{2\Omega}}|\psi_+\rangle\right]\nonumber\\
&\times e^{-i\varepsilon_et/\hbar},\label{Fe}
\end{align}
where $\Omega=\sqrt{(2\mu_BH_0/\hbar)^2+\omega_0^2}$,
\begin{eqnarray}
\varepsilon_g&=&\frac{\hbar\omega_0}{2}-\frac{\hbar\Omega}{2},\label{Eg}\\
\varepsilon_e&=&-\frac{\hbar\omega_0}{2}+\frac{\hbar\Omega}{2},\label{Ee}
\end{eqnarray}
are the Floquet (quasi)energies of these states, the indices $g$ and $e$ mark the ground and excited eigenstates, respectively,
and $|\psi_\pm\rangle$ are the eigenspinors of the operator $\sigma_z$, which satisfy the equation $\sigma_z|\psi_\pm\rangle=\pm|\psi_\pm\rangle$ and correspond to the mutually opposite directions of the spin along the $z$ axis (see Fig.~1a). Since the Floquet states (\ref{Fg})--(\ref{Fe}) with the energies (\ref{Eg})--(\ref{Ee}) are the exact solutions of the Floquet problem with the Hamiltonian (\ref{H}) and the field (\ref{H0}), they can be easily verified by direct substitution into the Schr\"odinger equation, $i\hbar\partial_t|\psi_{g,e}\rangle=\hat{\cal H}_0|\psi_{g,e}\rangle$.
\begin{figure}[h!]
\centering\includegraphics[width=0.8\columnwidth]{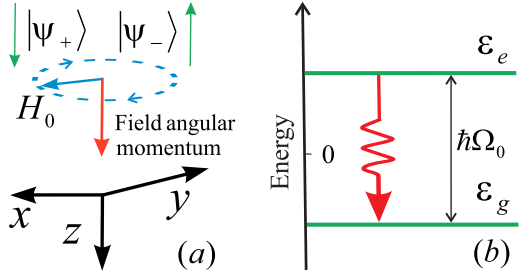}
\caption{(a) The basis spin states, $|\psi_+\rangle$ and $|\psi_-\rangle$, in the circularly polarized electromagnetic field with the magnetic field amplitude $H_0$, where the vertical arrow marks direction of the field angular momentum; (b) The energies of the Floquet eigenstates, $\varepsilon_g$ and $\varepsilon_e$, where the wave arrow marks the optical transition between these states.} \label{fig1}
\end{figure}

It follows from Eqs.~(\ref{Eg}) and (\ref{Ee}) that the dressing field (\ref{H0}) lifts the spin degeneracy (see Fig.~1b) and the field-induced spin splitting of the Floquet states (\ref{Fg})--(\ref{Fe}) is
\begin{equation}\label{S}
\hbar\Omega_0=\varepsilon_e-\varepsilon_g=\hbar(\Omega-\omega_0).
\end{equation}
The split Floquet states (\ref{Fg}) and (\ref{Fe}) are exact under the assumption of stability of them. However, the spin interaction with the vacuum photon mode of the frequency $\Omega_0$ can lead to the photon emission at this frequency, what corresponds to the optical transition marked by the wave arrow in Fig.~1b. As a consequence, decay of the excited state (\ref{Fe}) appears. Since the interaction with the photon vacuum can be described accurately only within the quantum field approach, one has to rewrite the Hamiltonian of the considered spin-field system in the quantized form in order to take into account this decay. Within the conventional quantum field theory (see, e.g., Ref.~\onlinecite{Scully_book}), the total spin-photon Hamiltonian reads
\begin{equation}\label{Hsf}
\hat{\cal H}=\hbar\omega_0\hat{a}_0^\dagger\hat{a}_0+\mu_B{\bm{\sigma}}\hat{\mathbf{H}}+\sum_{\mathbf{k}}\hbar\omega_{\mathbf{k}}\hat{a}_{\mathbf{k}}^\dagger\hat{a}_{\mathbf{k}}+\mu_B{\bm{\sigma}}\sum_{\mathbf{k}}\hat{\mathbf{H}}_{\mathbf{k}},
\end{equation}
where $\hat{a}_0^\dagger\,\,(\hat{a}_0)$ is the operator of photon creation (annihilation) for the dressing field (\ref{H0}) written in the Schr\"odinger picture (the representation of occupation numbers), $\hat{a}_{\mathbf{k}}^\dagger\,\,(\hat{a}_{\mathbf{k}})$ is the operator of photon creation (annihilation) for vacuum photons with the frequency $\omega_{\mathbf{k}}$ and the wave vector $\mathbf{k}$ in the same representation, $\hat{\mathbf{H}}=\sqrt{2\pi\hbar\omega_0/V}(\mathbf{e}_{0}\hat{a}_0+\mathbf{e}^\ast_{0}\hat{a}_0^\dagger)$ is the operator of the quantized dressing field (\ref{H0}), $\hat{\mathbf{H}}_{\mathbf{k}}=\sqrt{2\pi\hbar\omega_{{\mathbf{k}}}/V}(\mathbf{e}_{{\mathbf{k}}}\hat{a}_{\mathbf{k}}+\mathbf{e}^\ast_{{\mathbf{k}}}\hat{a}_{\mathbf{k}}^\dagger)$ is the operator of the vacuum field with the polarization $\mathbf{e}_{{\mathbf{k}}}$, and $V$ is the quantization volume. The first and third terms of the Hamiltonian (\ref{Hsf}) describe the energy of the dressing field (\ref{H0}) and the vacuum field, respectively, whereas the second and fourth terms describe the spin interaction with these fields.

Applying the quantum-field approach~\cite{Kibis_2010} developed to describe the pseudospin electron states in irradiated graphene to the considered spin problem, let us introduce the joint spin-photon space,
\begin{equation}\label{eps}
|\psi_\pm,N,n_\mathbf{k}\rangle=|\psi_\pm\rangle\otimes|N\rangle\otimes|n_\mathbf{k}\rangle,
\end{equation}
which describes the spin in the state $|\psi_\pm\rangle$, the quantized dressing field (\ref{H0}) in the state with the photon occupation number $N=1,2,3,...$, and the vacuum field in the state with the photon occupation number $n_\mathbf{k}=0,1$. Certainly, the basic states of this space meet the orthonormal conditions,
\begin{eqnarray}
\langle\psi_\pm,N,n_\mathbf{k}|\psi_\pm,N^\prime,n^\prime_\mathbf{k}\rangle&=&\delta_{N,N^\prime}\delta_{n_\mathbf{k},n^\prime_\mathbf{k}},\label{onc1}\\
\langle\psi_\pm,N,n_\mathbf{k}|\psi_\mp,N,n_\mathbf{k}\rangle&=&0.\label{onc2}
\end{eqnarray}
In the following, let us consider the spin interaction with the vacuum [the last term of the Hamiltonian (\ref{Hsf})] as a perturbation. Then the exact eigenstates of the unperturbed Hamiltonian [the first three terms of the Hamiltonian (\ref{Hsf})] can be written as
\begin{eqnarray}
|g, 1_\mathbf{k}\rangle&=&e^{-i\varepsilon_gt/\hbar}e^{-i\omega_\mathbf{k}t}\left[\sqrt{\frac{\Omega+\omega_0}{2\Omega}}|\psi_+,N_0,1_\mathbf{k}\rangle\rangle\right.\nonumber\\
&-&\left.\sqrt{\frac{\Omega-\omega_0}{2\Omega}}|\psi_-,N_0+1,1_\mathbf{k}\rangle\right]e^{-iN_0\omega_0t},\label{FFg}\label{FFg}\\
|e,0_\mathbf{k}\rangle&=&e^{-i\varepsilon_et/\hbar}\left[\sqrt{\frac{\Omega+\omega_0}{2\Omega}}|\psi_-,N_0,0_\mathbf{k}\right.\rangle\nonumber\\
&+&\left.\sqrt{\frac{\Omega-\omega_0}{2\Omega}}|\psi_+,N_0-1,0_\mathbf{k}\rangle\right]e^{-iN_0\omega_0t},\label{FFe}
\end{eqnarray}
where $N_0\gg1$ is the photon occupation number of the quantized dressing field (\ref{H0}). These eigenstates can be easily verified by direct substitution of them into the Schr\"odinger equation with the unperturbed Hamiltonian, keeping in mind that the classical field amplitude is $H_0=\sqrt{4\pi N_0\hbar\omega_0/V}$.

Since the quantized field (\ref{H0}) remains classically strong, the quantum description of the field and the classical one lead to the physically equal results. Particularly, the quantized spin-photon states (\ref{FFg})--(\ref{FFe}) are physically equal to the Floquet spin states (\ref{Fg})--(\ref{Fe}) arisen from the classical description of the same field. The only difference is the emitted photon with the frequency $\omega_{\mathbf{k}}$, which is added to the spin-photon state (\ref{FFg}). Namely, the state (\ref{FFe}) corresponds to the spin subsystem in the excited state (\ref{Ee}), whereas the state (\ref{FFg}) describes the spin subsystem in the ground state (\ref{Eg}) with the emitted photon of the frequency $\omega_{\mathbf{k}}$. Then the dynamics of the optical transition marked by the wave arrow in Fig.~1b can be described by the linear combination of the spin-photon eigenstates (\ref{FFg}) and (\ref{FFe}),
\begin{equation}\label{FFFg}
|\psi(t)\rangle=c_e(t)|e,0_\mathbf{k}\rangle+\sum_\mathbf{k}c_{g\mathbf{k}}(t)|g,1_\mathbf{k}\rangle,
\end{equation}
under the initial conditions $c_e(0)=1$ and $c_g(0)=0$. Substituting the spin-photon state (\ref{FFFg}) into the Schr\"odinger equation with the total Hamiltonian (\ref{Hsf}), we arrive at the quantum dynamics equations for the probability amplitudes $c_g(t)$ and $c_e(t)$,
\begin{eqnarray}
i\hbar \dot{c}_e(t)&=&\sum_\mathbf{k}U^\ast_\mathbf{k}e^{i(\Omega_0-\omega_\mathbf{k})t}{c}_{g\mathbf{k}}(t)\label{qd1}\\
i\hbar\dot{c}_{g\mathbf{k}}(t)&=&U_\mathbf{k}e^{-i(\Omega_0-\omega_\mathbf{k})t}{c}_e(t)\label{qd2},
\end{eqnarray}
where
\begin{eqnarray}\label{V}
{U}_\mathbf{k}&=&\langle g,1_\mathbf{k}|\mu_B\bm{\sigma}\hat{\mathbf{H}}|e,0_\mathbf{k}\rangle\nonumber\\
&=&
\frac{\mu_B(\Omega+\omega_0)}{2\Omega}\sqrt{\frac{2\pi\hbar\omega_\mathbf{k}}{V}}
\langle\psi_+|\bm{\sigma}\mathbf{e}_\mathbf{k}|\psi_-\rangle
\end{eqnarray}
is the matrix element of the spin interaction with the vacuum photon mode of the frequency $\omega_\mathbf{k}$. After integrating Eq.~(\ref{qd2}) and substituting it into Eq.~(\ref{qd1}), we obtain the equation,
\begin{equation}\label{die1}
\dot{c}_e(t)=-\sum_\mathbf{k}\frac{|U_\mathbf{k}|^2}{\hbar^2}\int_0^tdt^\prime e^{i(\Omega_0-\omega_\mathbf{k})(t-t^\prime)}c_{e}(t^\prime),
\end{equation}
which can be transformed by substituting Eq.~(\ref{V}) and replacing the summation over all vacuum photon modes $\mathbf{k}$ by the corresponding integrals. As a result, we arrive at the differential-integral equation,
\begin{eqnarray}\label{die2}
\dot{c}_e(t)&=&-\frac{\mu_B^2}{3\pi\hbar c^3}\left[\frac{\Omega+\omega_0}{\Omega}\right]^2\int_0^\infty d\omega_\mathbf{k}\,\omega_\mathbf{k}^3\nonumber\\
&\times&\int_0^tdt^\prime e^{i(\Omega_0-\omega_\mathbf{k})(t-t^\prime)}c_e(t^\prime),
\end{eqnarray}
which is still exact. Since the emitted photon frequency, $\omega_\mathbf{k}$, is centered about the transition frequency $\Omega_0$ marked in Fig.~1b, the quantity $\omega_\mathbf{k}^3$ varies little around $\omega_\mathbf{k}=\Omega_0$ for which the time integral in Eq.~(\ref{die2}) is not negligible. Therefore, one can replace $\omega_\mathbf{k}^3$ by $\Omega_0^3$  and the lower limit in the $\omega_\mathbf{k}$ integration by $-\infty$ (the Weisskopf-Wigner approximation~\cite{Scully_book}). Then Eq.~(\ref{die2}) yields
${c}_e(t)=e^{-\Gamma t/2}$, where the decay constant reads
\begin{equation}\label{G}
\Gamma=\frac{2\mu_B^2\Omega_0^3}{3\hbar c^3}\left[\frac{\Omega+\omega_0}{\Omega}\right]^2.
\end{equation}
It follows from Eq.~(\ref{FFFg}) that the photon emission is accompanied by the spin transition to the ground state (\ref{Fg}). Taking into account the normalization condition $|c_e|^2+\sum_\mathbf{k}|c_{g\mathbf{k}}|^2=1$, the probability of this process during time $t$ is
\begin{equation}\label{w0}
\sum_\mathbf{k}|c_{g\mathbf{k}}|^2=1-e^{-\Gamma t}.
\end{equation}

\section{Results and discussion}
The theory developed in Sec.~II describes the particular case of a two-level system with decay under a high-frequency driving field. It should be noted that similar two-level models are extensively used to analyze various atomic systems driven by the field (see, e. g., Refs.~\onlinecite{Scully_book,Meystre_book,Tannoudji_book}). To proceed, one need to extend the theory to the case of a pulsating field. As an example, let us consider the non-polarized spin system consisting of $n$ non-interacting spins, where $n/2$ spins are in the state $|\psi_+\rangle$ and $n/2$ spins are in the state $|\psi_-\rangle$. We aimed to find the spin polarization of the system under pulse of the field (\ref{H0}). For definiteness, we will restrict the following analysis by the case of the rectangular pulse, assuming the field (\ref{H0}) to be switched on at $t=0$ and switched off at $t=\tau_0$, where $\tau_0$ is the pulse duration (see Fig.~2a). Let a spin be initially in the state $|\psi_-\rangle$. Since the Floquet states (\ref{Fg}) and (\ref{Fe}) form the complete orthonormal basis for a spin in the field (\ref{H0}), the spin behaviour under the pulse can be described in the most general form by the spinor
\begin{equation}\label{Psi}
|\Psi(t)\rangle=\left\{\begin{array}{rl}
|\Psi_1(t)\rangle=|\psi_-\rangle,
&t<0\\\\
|\Psi_2(t)\rangle=a_e|\psi_e(t)\rangle+a_g|\psi_g(t)\rangle,
&0<t<\tau_0\\\\
|\Psi_3(t)\rangle=b_+|\psi_+\rangle+b_-|\psi_-\rangle,
&t>\tau_0
\end{array}.\right.
\end{equation}
Substituting Eqs.~(\ref{Fg})--(\ref{Fe}) into Eq.~(\ref{Psi}) and applying the continuity condition $|\Psi_1(0)\rangle=|\Psi_2(0)\rangle$, one can find the coefficients
$a_e=\sqrt{(\Omega_0+\omega_0)/2\Omega}$ and $a_g=-\sqrt{(\Omega_0-\omega_0)/2\Omega}$, which define the spinor $|\Psi_2(t)\rangle$. As expected, these coefficients satisfy the normalization condition $|a_e|^2+|a_g|^2=1$.
\begin{figure}[h!]
\centering\includegraphics[width=1.0\columnwidth]{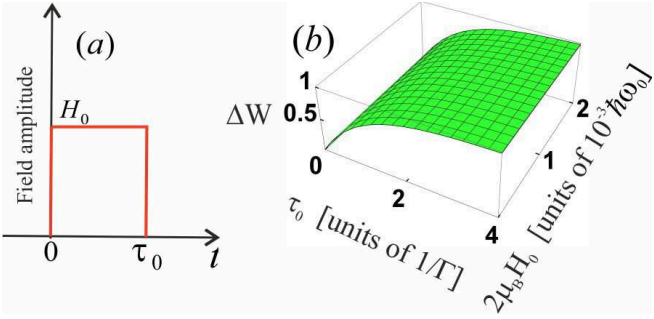}
\caption{(a) The rectangular pulse of the field with the amplitude $H_0$ and the duration $\tau_0$; (b) The spin-flip anisotropy probability, $\Delta W$, as a function of the field amplitude, $H_0$, and the pulse duration, $\tau_0$, for the photon energy $\hbar\omega_0=1$~eV.} \label{fig2}
\end{figure}

Taking into account the photon emission process discussed at the end of Sec.~II, there are the two physically different scenarios of the spin evolution during the pulse ($0<t<\tau_0$). In the first one, a photon is emitted by the spin, what is accompanied by the spin transition to the ground state $|\psi_g(t)\rangle$ defined by Eq.~(\ref{Fg}). It follows from Eqs.~(\ref{w0}) and (\ref{Psi}) that the probability of this scenario is $w=|a_e|^2(1-e^{-\Gamma\tau_0})$. Applying the continuity condition $|\psi_g(\tau_0)\rangle=|\Psi_3(\tau_0)\rangle$, one can find the coefficients $b_-$ and $b_+$ which define the spin state $|\Psi_3(t)\rangle$ after the pulse ($t>\tau_0$).  As a result, the probability of the spin flip process accompanied by the photon emission is $w|b_+|^2$. In the second scenario, the photon is not emitted and the spin remains in the state $|\Psi_2(t)\rangle$. The probability of this scenario is $1-w$. Applying the continuity condition $|\Psi_2(\tau_0)\rangle=|\Psi_3(\tau_0)\rangle$, one can find the coefficients $b_-$ and $b_+$ for this case. Correspondingly, the probability of the spin flip process without the photon emission is $(1-w)|b_+|^2$. The total probability of the spin flip induced by the pulse is the sum of the spin-flip probabilities found within these two scenarios. Going in the same way, one can find also the spin-flip probability if the spin is initially in the state $|\psi_+\rangle$. As a result, we arrive at the total probability of the spin flip induced by the pulse,
\begin{align}\label{W}
&W_\mp=\left(\frac{\Omega\pm\omega_0}{2\Omega}\right)^2\left(1-e^{-\Gamma\tau_0}\right)
+4\left(\frac{\mu_BH_0}{\hbar\Omega}\right)\sin^2\left(\frac{\Omega \tau_0}{2}\right)\nonumber\\
&\times\left[1-\left(\frac{\Omega\pm\omega_0}{2\Omega}\right)\left(1-e^{-\Gamma\tau_0}\right)\right],
\end{align}
where the signs ``$\mp$'' indicate the initial spin state, $|\psi_-\rangle$ or $|\psi_+\rangle$. It follows from Eq.~(\ref{W}) that the spin flip probability depends on the initial spin direction. Paricularly, the spin-flip anisotropy probability is
\begin{eqnarray}\label{DW}
\Delta W&=&W_--W_+=\frac{\omega_0}{\Omega}\left[1-
\left(\frac{2\mu_BH_0}{\hbar\Omega}\right)\sin^2\left(\frac{\Omega \tau_0}{2}\right)\right]\nonumber\\
&\times&\left(1-e^{-\Gamma\tau_0}\right).
\end{eqnarray}
As a consequence, the non-polarized spin system gets the pulse-induced spin directed along the field angular momentum,
$S_z=(n\hbar/2)\Delta W$. Thus, the light-induced magnetization of the system appears. It should be noted that the magnetization reverses its direction when reversing helicity of the field. Although the field (\ref{H0}) is assumed to be clockwise polarized, the reversed case of counter-clockwise polarization corresponds to the formal replacing $\omega_0\rightarrow-\omega_0$ in all equations. As expected, such a replacing turns the probability $W_+$  into the probability $W_-$ and vice versa [see Eq.~(\ref{W})].
It should be noted also that the models developed before to describe the light-induced magnetization are based on the interaction processes in multi-spin systems ~\cite{Kirilyuk_2010,Kimel_2015}, whereas the present theory results in the magnetization effect even within the simplest single-spin model.

It should be stressed that the photon emission at the frequency $\Omega_0$ is crucial for this magnetization effect. Indeed, it follows from Eqs.~(\ref{W}) and (\ref{DW}) that the asymmetry of the spin-flip process disappears ($\Delta W=0$) if the decay $\Gamma$ is zero. Correspondingly, the pulse-induced magnetization of the system also disappears if the photon emission is neglected. Therefore, the pulse-induced magnetization should be considered as a two-stage process. At the first stage, the pulse induces the spin splitting (\ref{S}). Physically, this splitting appears since a circularly polarized electromagnetic field acts similarly to a stationary  magnetic field. Indeed, both a magnetic field and a circularly polarized field breaks the time-reversal symmetry since it turns a clockwise-polarized field to a counter-clockwise polarized one and vice versa. As a consequence, a circularly polarized field can split electron states with mutually opposite orientations of both orbital angular momentum~\cite{Kibis_2011,Kibis_2021} and spin. At the second stage of the process, the photon emission at the splitting frequency occurs, what results in the discussed spin-flip anisotropy. It should be noted also that the physically similar mechanism of spin polarization under a rotating magnetic field takes place if the spin decay is induced by the presence of a stochastic magnetic field~\cite{Smirnov_1999}.

The dependence of the spin-flip anisotropy probability (\ref{DW}) on the field amplitude $H_0$ and the pulse duration $\tau_0$ is plotted in Fig.~2b for the field (\ref{H0}) with the photon energy $\hbar\omega_0=1$~eV. It follows from Eq.~(\ref{DW}), particularly, that the anisotropy strongly depends on the decay $\Gamma$. Within the developed single-spin theory, only the radiative process contributes to the spin decay (\ref{G}). However, there are the various non-radiative spin relaxation mechanisms originated from the spin-orbit and spin-spin interactions, which should also be taken into account in actual spin systems. Moreover, contribution of these mechanisms to the effective spin decay $\Gamma$ in solids can much exceed the relatively small radiative contribution (\ref{G}). Nevertheless, the present single-spin theory can be applied to complex spin systems as well. To describe such systems, the decay $\Gamma$ in the key equations (\ref{W}) and (\ref{DW}) should be replaced by $1/\tau_s$, where the spin relaxation time $\tau_s$ can be considered as a phenomenological parameter defined experimentally.

Finally, let us indicate the applicability limits of the present theory which is correct under assumption of $\omega_0\tau_s\gg1$. In solids, the spin relaxation time lies in the broad range from $\tau_s\sim10^{-6}$~s for electrons localized at impurities to $\tau_s\sim10^{-11}$~s for conduction electrons in materials with the strong spin-orbit coupling. Therefore, the field frequencies $\omega_0$ starting from the microwave range look appropriate to detect the considered effects.

\section{Conclusion}
Applying the Floquet theory of periodically driven quantum systems, the effect of a pulse of a circularly polarized electromagnetic field on the single-spin dynamics is considered. As a main result, it is demonstrated that the pulse induces the spin polarization along angular momentum vector of the circularly polarized field. The polarization process consists of the two stages. At the first stage, the pulse induces the spin splitting. At the second stage, the photon emission at the splitting frequency occurs. A a consequence of the two-stage process, the field-induced spin polarization and the corresponding magnetization of the system under consideration appears. Such a magnetization effect can be observed for various spin systems in modern experiments.

\begin{acknowledgments}
The reported study was funded by the Russian Foundation for Basic Research (project 20-02-00084). I am grateful for the support of the Russian Ministry of Science and Higher Education (project FSUN-2020-0004). \end{acknowledgments}

\end{document}